\documentclass[aps,prb,preprint,showpacs,floatfix,superscriptaddress]
{revtex4-1}

\usepackage{graphicx}

\begin{document}

\title{Thermally populated intrinsic localized modes in 
pure alkali halide crystals}

\author{A. J. Sievers} 
\affiliation{Laboratory of Atomic and Solid State Physics, 
Cornell University Ithaca, NY 14853-2501}
\author{M. Sato}
\affiliation{Graduate School of Natural 
Science and Technology, Kanasawa University, Kanasawa, Ishikawa 
920-1192, Japan} 
\author{J. B. Page} 
\affiliation{Department of Physics, Arizona State University,
Tempe, AZ 85287-1504}
\author{T. R\"ossler}
\affiliation{Yingli Green Energy Europe GMBH,
Heimeranstr. 37, D 80339 Munich, Germany}

\date{\today}

\begin{abstract} The possibility of thermal excitation of intrinsic
localized modes (ILMs) arising from anharmonicity in ionic perfect
crystals is studied numerically for realistic model systems in one
and three dimensions. Implications are discussed for an interesting
high-temperature feature seen in earlier inelastic neutron scattering
experiments on single crystal NaI. The general conclusion is that ILM
formation energies are far too large for thermal excitation of ILMs to
account for the observed feature in a pure crystal.
\end{abstract}


\maketitle

\section{\label{sec:Intro}Introduction}

The discovery that some localized excitations in nonlinear
perfect lattices can be stabilized by lattice discreteness
has led to extensive studies of their characteristic features.
\cite{Dol86,ST88,Page90,FW98,LS99,CFK04,FG08} Experimental studies
have focused on intrinsic localized modes (ILMs) driven in the steady
state to overcome energy loss to the lattice. Since spin lattices
are intrinsically nonlinear, the CW driving of antiferromagnetic
instabilities has made it possible to create and destroy 1D
intrinsic localized spin wave modes in the layered antiferromagnet
(C$_2$H$_5$NH$_3$)$_2$CuCl$_4$.\cite{SS08,WSS05} Additional details on
the creation and destruction of ILMs have been studied experimentally
for macroscopic 1D E\&M transmission lines\cite{ETS08,EPS10} and
micromechanical arrays.\cite{SHS06,Sato11} The common thread between the
microscopic and macroscopic systems is that after a driven instability
generates many ILMs, a few become stabilized by locking to the driver
frequency. Hysteresis and switching of ILMs have been demonstrated for
both kinds of nonlinear lattices.

In a 2005 numerical study, Eleftheriou and Flach\cite{EF05} addressed 
the problem of whether or not ILMs might be observable in thermal 
equilibrium. They used a highly simplified monatomic 2D model square lattice 
having a single scalar degree of freedom at each site. The interactions 
comprised nearest-neighbor quadratic and on-site quartic terms. For 
a fixed temperature the system was thermalized, and molecular dynamics 
(MD) calculations were used to generate time-dependent ensemble-averaged 
amplitude autocorrelation functions which were Fourier transformed to 
produce power spectra. Clear signatures of ILMs proved elusive, but by 
cooling the edge of their 2D lattice (via dissipation terms) so as to 
remove running lattice modes, the authors were able to see the remaining 
stationary ILMs. They noted that it would be difficult to carry out a 
corresponding procedure in any realistic laboratory experiment. 

Thus it was somewhat surprising that in 2009, inelastic neutron scattering
measurements on NaI: 0.3 mole \% T$ \ell\,$I showed that at a temperature 
of about 570 K a localized vibrational excitation rapidly appears near 
the center of the large phonon gap between the transverse optic and 
acoustic branches, polarized along the [111] direction.\cite{Manley09} 
This observation was interpreted as the appearance of a thermally 
generated ILM. Subsequent measurements at still higher crystal 
temperatures showed even more complex dynamical behavior, interpreted 
as a coherent rearrangement of ILMs in the atomic lattice.\cite{MA11}

A recent series of numerical studies by Dmitriev and coworkers have 
considered the theoretical question of high temperature excitation of 
localized vibrations in alkali halide crystals.\cite{DMCw} Employing
some of the nonzero temperature MD methods of Ref.\ \onlinecite{EF05} 
but using realistic interaction potentials, they argue via simulations 
on a 2D nonlinear lattice model that the vibrational lifetime of the 
light atoms in a system with a large harmonic phonon gap grows with 
increasing temperature, and they interpret this as evidence of ILMs. 
Their calculations did not use the edge-cooling technique of 
Ref.\ \onlinecite{EF05} to isolate the thermally generated stationary ILMs.

Most recently, inelastic neutron scattering experiments on the vibrational
spectrum of NaI up to 700 K have been reported with no evidence for a high
temperature gap mode.\cite{Kempa13} Given the variety of theoretical 
and experimental results, there is value in reassessing the likelihood of an 
experimentally significant fraction of thermally populated ILMs in alkali 
halide crystals.

Because of the simplicity of the ionic bond in alkali halides, 
they were one of the first crystal types whose interatomic potentials
and vibrational dynamics were well studied.\cite{AM76} The result is
that the Born-Mayer-Coulomb shell model potentials yield accurate fits
to measured phonon dispersion curves.\cite{BSW84} Since the resulting
ionic potentials are quite harmonic, ILM production in these systems
requires large vibrational amplitudes in order for the nonlinearity to
produce dynamical localization. In this report we examine the formation
energy necessary to produce an ILM in the phonon gap of alkali halide
crystals, a topic essentially unexplored in the literature. We find that
the necessary formation energy is much too large to produce a measurable
concentration of ILMs below the pure ionic crystal melting point.

\section{\label{sec:Formenergy}Formation energy estimates for ILMs in
ionic crystals}

We have computed the formation energy for two ionic crystal models,
a 1D model for KI and a previously published 3D model\cite{KS97} for
NaI. The large mass difference between the positive and negative ions
in each of these lattices produces a large frequency gap between the
optic and acoustic harmonic phonon bands.

\subsection{\label{subsec:KI}KI}

Early theoretical work on ILMs focused on 1D models and is reviewed
in Ref.\ \onlinecite{SP95}. As discussed there, the use of realistic
interatomic full pair potentials $V(r)$ for ionic crystals, as opposed to
harmonic and anharmonic spring models, rules out the existence of ILMs
with frequencies above the maximum harmonic lattice frequency. This is
because of the rapid softening of the interatomic forces with separation
$r$. ILMs can readily form, however, in the phonon gaps of diatomic
lattices.

Here we report calculated formation energies for gap ILMs in KI using
interatomic Born-Mayer plus Coulomb (BMC) potentials. Specifically,
the model is a 1D lattice of alternating masses $m$ and $M$ moving
longitudinally, with nearest neighbors interacting via
\begin{equation}
\label{BMCpotential}
V_{mM}(r) = V_{Mm}(r) = \lambda e^{-r/\rho} -\frac{q^2}{r}, 
\end{equation}
while second neighbors interact via
pure Coulomb potentials 
\begin{equation} 
V_{mm}(r) = V_{MM}(r) = \frac{q^2}{r}.
\end{equation}
More distant neighbors are assumed to be noninteracting. 
Periodic boundary conditions (PBCs) are used, for a lattice of 
$N =$ 40 particles.

Our parameter values were determined by fitting measured harmonic phonon
dispersion curves \cite{Bilz79} along the [111] direction for KI. The
fit is excellent. Our model parameters are $m = 39.1$ amu, $M=$ 127 amu,
$\lambda=$ 2.57 $\times$ 10$^4$ eV, $\rho=$ 0.289 $\AA$, and $q=$ 0.90
e. Minimization of the total potential energy yields the static lattice
nearest-neighbor separation $a=$ 3.50 $\AA$. The lowest frequency of the
harmonic optic phonon band is at the zone boundary, with $\omega_{zbm}=
2.47 \times 10^{13}$ rad/sec, equivalent to 16.3 meV in energy units.

Using the rotating wave approximation (RWA) for the particles' time
dependence,\cite{Dol86,ST88,SP95,FW98,RP00} we obtained accurate
predictions for stationary ILM solutions of the classical equations of
motion. Briefly, one assumes that the position of atom $n$ is described by
\begin{equation}
\label{RWA}
r_n(t) = b_n +c_n \cos (\omega t) + r^0_n,
\end{equation}
where $r^0_n$ is the equilibrium position. Substitution
into the equations of motion, multiplication by either unity or
$\cos(\omega t)$, followed by an integration over a single period yields
a system of $2N$ coupled nonlinear time-independent equations for the
static and dynamic displacements $\{b_n\}$ and $\{c_n\}$. The equations
are solved numerically, and the predictions are checked by MD simulations. 
It is straightforward to add higher harmonic terms to Eq.\ (\ref{RWA}), 
but their contributions are typically down by an order of magnitude.

\begin{figure}[ht] 
\includegraphics[width=8cm]{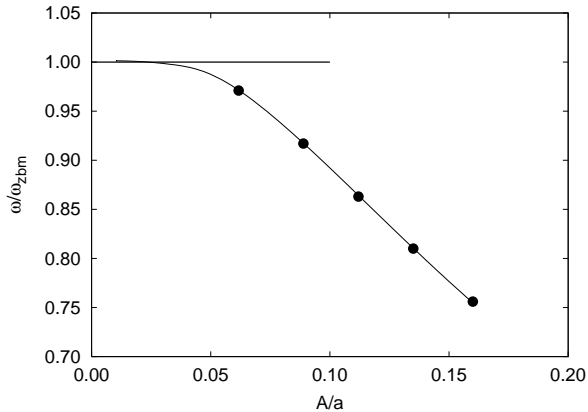}
\caption{\label{fig1} Frequency vs. amplitude for gap ILMs in the 1D model
for KI. The amplitudes $A$ are the dynamic displacements of the ILM's
central particle and are normalized to the nearest-neighbor distance
$a$. The frequencies are normalized to the harmonic phonon optic band
minimum $\omega_{zbm}$. In this normalization, the harmonic phonon gap
extends from 1. to 0.51. The circles denote the five ILMs for which
formation energies are computed.}
\end{figure}

\begin{figure}[ht] 
\includegraphics[width=8cm]{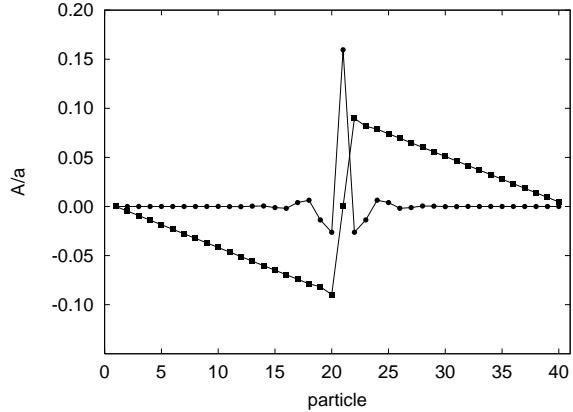}
\caption{\label{fig2} Displacement patterns for the ILM at
($\omega/\omega_{zbm},A/a) = (0.76,0.16)$ in the 40 particle 1D model
for KI. The dynamic and static displacements are denoted by circles
and squares, respectively. The actual displacements are longitudinal,
but are plotted vertically for clarity. The static displacements vanish
at the ends, due to the use of periodic boundary conditions.}
\end{figure}

Figure \ref{fig1} shows the predicted ILM frequency as a function of
the ILM's central particle dynamic displacement $A$, along with symbols
marking the ILMs whose formation energy were computed. The dynamic and
static displacement patterns for the most localized of these are shown
in Fig.\ \ref{fig2}. This ILM is represented by the rightmost circle
on the curve of Fig.\ \ref{fig1}, and as one moves to the left along
the curve, the ILM spatially broadens, eventually becoming the zone
boundary mode. We did not run formal stability calculations (i.e.,
Floquet) for the ILMs, but MD runs showed that our highest frequency
ILM ($\omega/\omega_{zbm} = 0.97$) persists for more than 900 periods,
while that at ($\omega/\omega_{zbm} = 0.92$) is numerically stable for
more than 200 periods. Our most localized ILM (at $\omega/\omega_{zbm} =
0.76$) is numerically stable for $\approx$ 50 periods, after which the
central particle displacement slowly spreads into the lattice. The MD
frequencies for the five ILMs in KI are within two to five percent of
their RWA predicted values as the frequencies descend into the gap.

The ILM classical energy $E_0$ at $T=0$ K is conveniently computed via
two single time-step MD runs. First, we sum the predicted dynamic and
static displacements at each site to obtain the ILM's initial displacement
pattern $\{r_n(0) = b_n + c_n + r^0_n\}$, for which the kinetic energy is
zero. Since the potential energy (PE) and forces are computed at every
time step in MD before the particles move, the first timestep provides
the full PE (including that from the static distortion) for the $t=0$
ILM configuration. Subtracting from this the similarly obtained PE for
the equilibrium configuration of the lattice $\{b_n = c_n =0\}$ then
yields the ILM's full classical energy.

\begin{figure}[ht] 
\includegraphics[width=8cm]{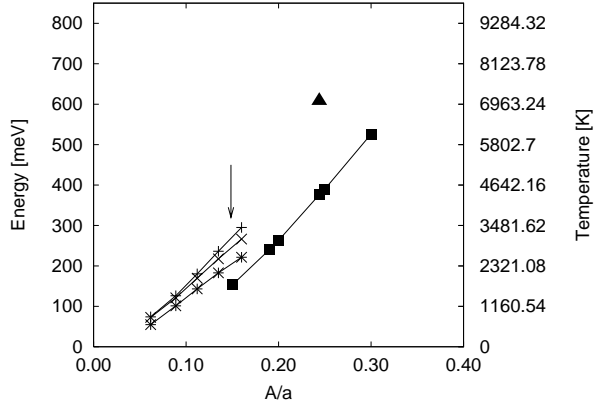}
\caption{\label{fig3} ILM energies $E_0$ versus ILM amplitude, computed
for the 1D KI model and the 3D NaI model.  For KI, the crosses are for
the full MD method described in Sec.\ \ref{subsec:KI}, and the $ \times $
and star symbols are for the two approximation methods introduced there,
namely RWA$_{KE}$ and RWA$_{KEcentral}$, respectively.  The squares give
NaI results computed using the RWA$_{KEcentral}$ approximation, while
the single triangle is the result of the better RWA$_{KE}$ approximation,
with the sum in Eq.\ (\ref{RWAenergy}) going over the ILM central particle
and two nearest neighbor shells. The energies $E_0$ are relative to the
$T=0K$ equilibrium configuration of the lattice. The vertical arrow is
explained in the text.}
\end{figure}

Applying the above procedure for the five ILMs indicated on the curve of
Fig.\ \ref{fig1}, we obtained the results shown by the crosses in Fig.\
\ref{fig3}. This figure also includes results for the 3D model of NaI,
which will be discussed later. Here we focus on the energies for KI,
shown in the lower left part of the figure. The results given by the
crosses were obtained using the single-step MD calculations described
above, and one sees that the energies are large, ranging from 70 to
300 meV. The temperature scale on the right-hand side of the figure
shows that the corresponding equivalent temperature $T=E_0/k_B$ of the
lowest-energy ILM, for ($\omega/\omega_{zbm}, A/a) = $(0.97,0.062),
is 858 K, near the KI melting point of 954 K. The other four energies
are well above that temperature.

In addition to the full $E_0$ results, Fig.\ \ref{fig3} gives two
sets of KI results which follow from simple approximations, based on
the assumed RWA time dependence of Eq.\ (\ref{RWA}). Within the RWA
the maximum kinetic energy of an ILM is given by the familiar harmonic
approximation result 
\begin{equation}
\label{RWAenergy} 
E_{RW\!\!A_{KE}} = \frac{\omega^2}{2} \Sigma_{n=1}^N m_n c_n^2, 
\end{equation}
where the sum is over the $N$ particles of the lattice. This 
approximation requires only the predicted frequencies 
and dynamic displacements. Despite the facts that an ILM has a 
static component $\{b_n\}$ which does not
contribute to the above formula and also that the RWA is itself not
exact, this approximation turns out to be quite good, as is seen by the
$\times$ symbols in Fig.\ \ref{fig3}. As one expects, the results of
this approximation become very close to the full $E_0$ results for our
smallest-amplitude ILM at $A/a$ = 0.062. An even more approximate result
is to include {\it only} the contribution from the ILM's central particle
($A^2$ term) in the sum of Eq.\ (\ref{RWAenergy}). The results for this
``RWA$_{KEcentral}$'' approximation are given by the stars in Fig.\
\ref{fig3}.

The more exact full energies $E_0$ and their two RWA approximation
counterparts in Fig.\ \ref{fig3} are seen to be quite close. More
importantly, the results from all three methods cover nearly the same
equivalent temperature range, much higher than the 570 K temperature at
which the anomalous gap mode feature appears in the inelastic neutron
scattering data of Ref.\ \onlinecite{Manley09}.

\subsection{\label{subsec:NaI}NaI}

To estimate the $T=0$ K classical energy for an ILM in 3D NaI, we
draw on the results presented for the 1D KI case treated above. Reference
\onlinecite{KS97} summarizes the 3D model used for NaI, and additional
details are given in Ref.\ \onlinecite{KSC98}. Briefly, a simulated
annealing technique was combined with a rigid-ion two-body potential
model of NaI to predict the properties of its ILMs for two different
ionic lattice structures, namely zincblende and fcc. The predictions
were verified with MD. The results are summarized in Fig.\ 4 of Ref.\
\onlinecite{KS97}, which shows the ILM frequency versus amplitude curves
for the two structures. As for the previous section, the amplitude is the
predicted dynamic displacement of the ILM's central particle. The general
behavior is similar for both lattice structures in that with increasing
mode amplitude the ILM frequency drops farther into the harmonic phonon
gap between the TO and TA branches.  The difference between the curves for
the two lattices is that for the zincblende structure the ILM frequency
decreases much more rapidly with increasing ILM amplitude. For 3D NaI
an amplitude threshold is observed at about $A/a = 0.1$ (a similar,
but smaller, threshold is seen for 1D KI in Fig.\ \ref{fig1}). For NaI
it was found that at a relative amplitude of 0.244 the ILM normalized
frequency is $\omega/\omega_{zbm}=0.95$, where $\omega_{zbm}$ is the
frequency of the harmonic zone boundary TO mode in the [111] direction.
For this ILM the predicted static and dynamic displacements are given
for the central particle and four shells of neighbors in Table I of Ref.\
\onlinecite{KS97}. This ILM had the longest lifetime, $\approx$ 200--250
periods, via MD runs. For an ILM frequency of $\omega/\omega_{zbm} =
0.90$, the lifetime decreased to $\approx$ 100 periods. These numerical
lifetimes are qualitatively similar to those given above for the 1D KI
model over the same range of $\omega/\omega_{zbm}$. The NaI predicted
and MD frequencies were plotted out to a relative amplitude $A/a$ of
0.3 in Ref.\ \onlinecite{KS97}, but the ILM energies were not recorded.

The solid squares in Fig.\ \ref{fig3} give energies for 3D NaI ILMs
computed using the ``RWA$_{KEcentral}$'' approximation defined below
Eq.\ (\ref{RWAenergy}) and which we have seen works quite well for 1D
KI. The ILM central particle amplitudes needed for this approximation
were obtained from Fig.\ 4 of Ref.\ \onlinecite{KS97}.  Again following
the results for KI, an improved estimate can be obtained from Eq.\
(\ref{RWAenergy}) by including more than just the ILM's central particle
in the sum. The triangle in Fig.\ \ref{fig3} gives the result for the
ILM having relative amplitude ($A/a$=0.244). The sum included the central
particle and its two nearest shells of neighbors, with the displacements
taken from Table I of Ref.\ \onlinecite{KS97}. Similar to the 1D KI
calculation, the complete full energy for this amplitude should be
somewhat larger, and the energy curve should grow rapidly with increasing
amplitude, just as for the central-particle estimate. One major difference
between our energy results for 1D and 3D is highlighted by the arrow in
Fig.\ \ref{fig3}, which shows the position of the 1D KI relative amplitude
corresponding to the relative frequency $\omega/\omega_{zbm}=0.78$ of
the experimentally measured NaI gap mode.  Within the 3D ILM model for
NaI, this relative frequency of 0.78 shifts the arrow into a nonphysical
region where the relative amplitude would be larger than $A/a =$ 0.4. This
would correspond to a huge ILM energy (note that simply extrapolating
the solid squares to $A/a$=0.4 would give a lower bound of 9000 K for
the ILM's energy equivalent temperature).\cite{symmetryremark}

We have seen that the classical energy $E_0$ needed to create a mid-gap
ILM in our 40 particle 1D KI model is of order 300 meV. We emphasize
that this energy is relative to that for the $T=0$ K classical lattice
equilibrium configuration.  As pointed out earlier, with decreasing
amplitude the ILM spatially broadens, eventually becoming the harmonic
zone boundary mode. At a high temperature, viz. 570 K, and with no ILM
present, the harmonic zone boundary mode has energy $k_B T = 49$ meV.
As a first approximation then, we imagine the ILM's ``parent'' zone
boundary mode converting to the ILM, and thereby take the ILM formation
energy to be $E_F = E_0 - k_B T = 295 - 49 = 246$ meV. For the 3D NaI
model the situation is different since the ILM's harmonic ``parent''
mode is the zbm at the L point along the [111] direction in the BZ. Since
there are four symmetry-equivalent L points which can convert to the
ILM, we take the formation energy in NaI to be $E_0 - 4 K_B T$.  The ILM
formation energy results are summarized in Table \ref{table1}.

\begin{table}[h]
\caption{\label{table1} ILM formation energies for the
1D KI model and the 3D NaI model, as listed in the first column. The
second column gives the ILM frequency in units of $\omega_{zbm}$,
the third column gives the corresponding dynamic amplitude of the ILM
central particle in units of the nearest-neighbor distance $a$. The
fourth column gives the calculated ILM energies $E_0$ plotted in Fig.\
\ref{fig3}, and column five gives our estimates of the ILM formation
energies for $T = 570$ K, using $E_F = E_0 - k_B T$ for KI and $E_F =
E_0 - 4 k_B T$ for NaI, as discussed in the text.}
\begin{ruledtabular}
\begin{tabular}{ccccc}
\textrm{Model} & $\omega/\omega_{zbm}$ & $A/a$ & $E_0$ & $E_F (570 K)$ \\ 
&&& \textrm{(meV)} & \textrm{(meV)}\\ 
\colrule
\textrm{KI 1D} & 0.971 & 0.0617 & 74. & 25.\\ 
\textrm{KI 1D} & 0.917 & 0.0889 & 126. & 77.\\
\textrm{KI 1D} & 0.863 & 0.112 & 180. & 131.\\
\textrm{KI 1D} & 0.810 & 0.135 & 236. & 187.\\ 
\textrm{KI 1D} & 0.756 & 0.160 & 295. & 246.\\ 
\textrm{NaI 3D} & 0.950 &0.244 & 608. & 412.\\
\end{tabular} 
\end{ruledtabular} 
\end{table}

In the configurational entropy model, the site-occupancy probability for a
localized lattice excitation, be it a vacancy or ILM, is
\begin{equation}
\label{probs} 
p(T)=\frac{1}{e^{T_F/T} + 1} 
\end{equation}
where $T_F$ is the formation temperature. Since our approximate ILM 
formation energy expressions above include small temperature corrections 
to $E_0$, we see that the corresponding formation temperatures are of 
the form $T_F(T) = E_0 /k_B - cT$, where the constant $c=$ 1 or 4 for 
the 1D and 3D models, respectively. For NaI at 570 K and using the 
value of $E_0$ given in Table \ref{table1}, we obtain $T_F=$ 4776 K 
and $p= 2.3 \times 10^{-4}$. At the NaI melting point of 934 K, these 
results become $T_F=$ 3328 K and $p= 2.8 \times 10^{-2}$.

\section{\label{sec:RelStrength}Relative strength of the local mode to
the zone boundary optic mode}

The idea here is to start from a 1D harmonic toy model with parameters
chosen so that the dispersion curve roughly represents that for
NaI in the [111] direction. The ILM is represented by a simple
force constant harmonic defect mode centered on a Na atom whose two
nearest-neighbor springs are weakened enough to produce a gap mode at
0.78 times the harmonic zone boundary frequency $\omega_{zbm}$, with
the remaining springs unchanged.  Denote the mode Na*. The properties
of the dispersion curves, especially near the zone boundaries are then
monitored numerically as the impurity concentration is increased. Our
aim is to tie the effects on the mode spectrum directly to the number
of gap modes without introducing the intermediate step of temperature.
The final step is to track the strengths of the zone boundary mode and
the gap mode when a random concentration of Na* is present and compare
the results with the experimental relative strengths shown in Fig. 2 of
Ref.\ \onlinecite{Manley09}.

\begin{figure}[ht] 
\includegraphics[width=8cm]{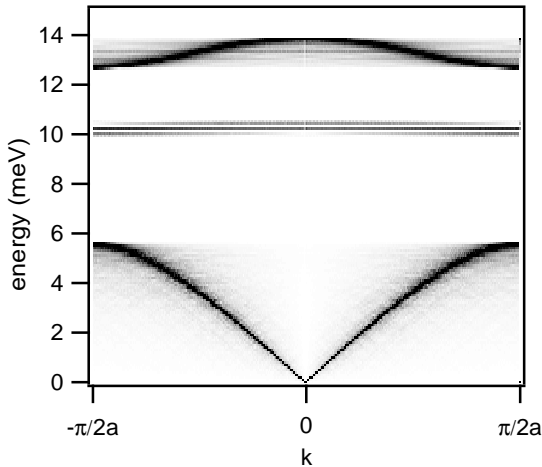}
\caption{\label{fig4} Numerical dispersion curves for the 1000 particle 1D
harmonic model NaI lattice with a 14\% concentration of Na* impurities.}
\end{figure}

To compare the strength of the harmonic gap mode to that of the zone 
boundary mode, we first obtained the normalized harmonic mode eigenvectors 
for the 1D model NaI host lattice, with 1000 particles. The eigenvectors
were Fourier transformed to obtain $\omega(k)$. Next the force constant
impurity was introduced, to reproduce the frequency of the assumed
localized gap mode, and this was followed by a series of calculations
of the normal modes for different random impurity concentrations. The
harmonic eigenvectors for the defect lattice were Fourier transformed and
their FT amplitudes were used to generate a perturbed ``dispersion curve''
as follows: for a mode of frequency $\omega$, the complex square amplitude
of its FT at $k$ was added to a frequency histogram bin ($\omega,
\omega+d\omega$). The square root of sum of all such contributions was
then plotted vs $k$, with the results shown in Fig.\ \ref{fig4}. One
sees that the resulting curve is smeared out horizontally due to the
fact that many of the modes are localized, so that their frequencies no
longer correspond to a single $k$ point. This feature is most pronounced
in the gap frequency region as expected, but it is also evident in the
optic branch as the impurity concentration increases. Figure \ref{fig4}
is for 70 Na* particles (14\% concentration).

\begin{figure}[ht]
\includegraphics[width=8cm]{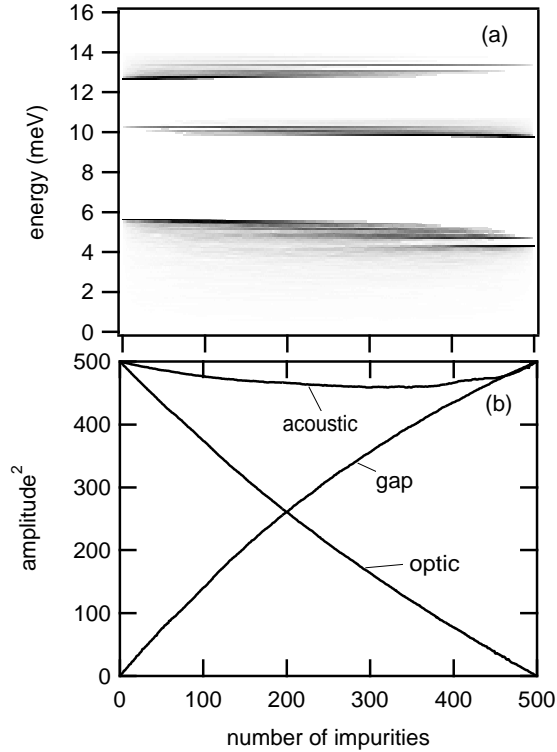}
\caption{\label{fig5} (a) Energy dependences of optic and acoustic zone
boundary modes and the gap mode for a force constant defect in the 1D
harmonic model diatomic lattice vs the number of Na* impurities. (b)
Integrated strengths at the zone boundary in each of the three energy
regions vs. impurity number.}
\end{figure}

Panel (a) of Fig.\ \ref{fig5} shows our resulting energy spectrum
near the host lattice zone boundary as a function of Na* impurity
concentration. The figure covers the entire concentration range from
zero to 500 impurities in our 1000 particle diatomic lattice, and two
important features are evident: (i) the gap region is well defined over
the whole impurity range and (ii) the optic branch loses strength as the
strength of the gap mode increases. We are particularly interested in the
lower concentration range. To examine the energy dependence of our results
more quantitatively, panel (b) of Fig.\ \ref{fig5} plots the integrated
strength in each of three energy regions versus concentration. Two key
features are evident: (i) The zone boundary strength coming from the
acoustic branch is fairly constant over the entire concentration range
and (ii) the strengths of the gap band and zone boundary optic band are
interconnected. Note that these two strengths are equal at an impurity
number of 200, corresponding to about 40\% Na*. The gap mode strength is
about half that of the optic mode when the number of impurities is 126,
corresponding to a concentration of about 25\%.

\begin{figure}[ht] 
\includegraphics[width=8cm]{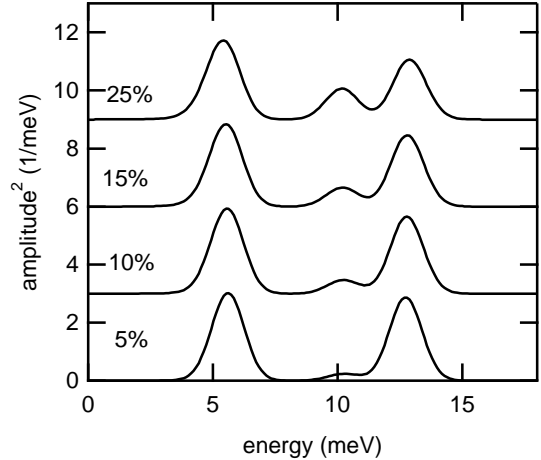}
\caption{\label{fig6} Slices of the zone boundary data of panel (a) of
Fig.\ \ref{fig5} examined with a Gaussian resolution function with 1.5
meV FWHM. The impurity concentrations are given in the figure, and the
area under each trace is normalized to unity. The curves are shifted
for clarity.}
\end{figure}

Because of the interconnection between the gap mode strength and the
optic band strength at the zone boundary, these mode-counting results
strongly suggest that the high temperature concentration of gap modes
observed in the neutron scattering experiment is large, rather than
small as was previously assumed.\cite{Manley09} To double check this
conclusion we took a Gaussian resolution function with 1.5 meV FWHM and
scanned it through the spectra shown in panel (a) of Fig.\ \ref{fig5} for
different concentrations. The results are given in Fig.\ \ref{fig6}. Note
that the gap mode strength is $\approx$ 1/2 of the optic zone boundary
mode strength, as represented by 25\% in Fig.\ \ref{fig6}. Clearly
the impurity concentration producing the gap modes must be comparable
with the concentration of normal Na ions producing the optic modes. In
addition the low resolution representation of the zone boundary mode
spectrum in Fig.\ \ref{fig6} shows that the gap modes could be ``seen''
in either the 10 or 15\% concentration spectra.

So far in this section, we have not invoked the concept of temperature. If
we assume that the concentration of Na* is due to a thermally activated
process with activation energy $k_B T_F$, the configurational entropy
model for vacancy production [Eq.(\ref{probs})] would apply. What is the
activation energy needed to produce a 25\% concentration of impurities at
about 570 K? From our present 1D zone boundary mode counting calculation,
the gap mode strength at 25\% impurities in the lattice is 1/3 and that
of the optic branch is 2/3. Hence 25\% corresponds to a probability of
0.33, and according to Eq.\ (\ref{probs}) for $T=570$ K we have $T_F=404$
K. Note that for this small energy barrier, the room temperature impurity
concentration would be $\approx$ 20\%, but no such evidence of gap mode
scattering is found at either 300 or 473 K in the neutron scattering
data.\cite{Manley09} In the experiment, the gap modes ``seem'' to appear
suddenly in that temperature range, with a 25\% concentration.

\section{\label{sec:DandC}Discussion and Conclusions}

How does the observed high temperature gap mode feature in NaI compare
with the work reported by others for NaCl type crystals? In the third
paper of Ref.\ \onlinecite{DMCw}, Kistanov and Dmitriev conclude from MD
calculations of the density of phonon states at elevated temperatures that
ILMs (discrete breathers) do exist for crystals with the NaCl structure
that have a gap between their optic and acoustic phonon branches. Our
inspection of their Fig.\ 3(d), which they use to support their claim,
indicates that what they have found at high temperatures is not due to
ILMs but is instead the modulation of the optical density of states by the
low-lying acoustic density of states due to the nonlinearity. We note that
the widths of the sum and difference frequency sidebands, as measured from
the peak in the optical density of states, each correspond to the width
of the acoustic density of states in the same figure, as would be expected
from this interpretation. There is no evidence of ILMs in this spectrum.

On the experimental side, the inelastic neutron scattering results
for NaI between 300 K and 700 K at 100 K intervals, presented in Ref.\
\onlinecite{Kempa13}, failed to show  any evidence of a gap mode at 600 K
at the zone boundary L point for energy spectra taken at ${\bf Q}=(2.5,
2.5, 1.5)$. (Measurements were not made at 555 K.)  The authors did
observe the gap region between the optic and acoustic branches to fill
in at high temperatures as would be expected from the modulation of the
optical density of states by the acoustic modes, as mentioned in the
preceding paragraph.

What we have described in this paper is consistent with these negative
findings.  For the 1D Born-Mayer-Coulomb model of KI, we computed
the ILM energies $E_0$ by three methods. Two of them are simple
approximations that require knowledge of only the ILM's frequency
and dynamic displacements, which are straightforward to predict in the
RWA. The third and most exact method is a direct calculation of the $t=0$
classical potential energy at the first step of an MD run that starts from
the RWA-predicted full ILM displacement pattern, static plus dynamic. For
ILMs in KI, the three methods were found to be in good agreement, with
the more exact method giving an energy $E_0$ of about 300 meV for an ILM
whose frequency is well within the harmonic phonon gap ($\omega = 0.76
\omega_{zbm}$). Of the two approximate methods, the roughest requires
the predicted frequency and dynamic displacement of only the central
particle, and for the same ILM in KI it yields about 220 meV, about 25\%
below the more exact value.

For NaI, we drew on the previously published 3D rigid ion model ILM
results of Ref.\ \onlinecite{KS97}.  Using that paper's predicted
ILM frequencies and central particle dynamic displacement in our
roughest approximation method for an ILM near the top of the gap at
$\omega/\omega_{zbm} = 0.95$, we estimate the energy $E_0$ to be near
400 meV. Using the better of the two approximation methods raises this
figure to just above 600 meV. But this ILM is near the top of the gap,
and we argued in Sec.\ \ref{subsec:NaI} that for an ILM deeper in the 
NaI phonon gap, the energy would be of the order of 1000 meV. This value
is comparable to that measured for the formation of vacancies in NaI
(1840 meV).\cite{KT70} Such large energies predict that both ILM and
vacancy concentrations would necessarily be very small, even near the
NaI melting temperature of 934 K. At the same time, our calculation in
Sec.\ \ref{sec:RelStrength} of the relative strength of a hypothetical
harmonic defect mode to that of the harmonic zone boundary mode in a 1D
harmonic diatomic crystal would require a defect concentration of about
25\% to match the experimental inelastic neutron scattering results at
570 K. We conclude that thermally excited ILMs cannot be the source of
the high temperature local mode observed in the phonon gap of NaI.

More generally, the MD simulations of Ref.\ \onlinecite{DMCw}, as described
earlier, show that the modulation spectrum of the optic modes produced
by the acoustic modes can be expected to have an undesirable experimental
consequence. In a vibrational lattice with wide harmonic frequency gaps, this 
process will necessarily mask the experimental observation of the low
concentration of ILMs lurking there. To expose ILMs experimentally in atomic
lattices it appears necessary to lower the temperature and externally drive
the lattice.\cite{RP00,RP95}

\begin{acknowledgments} A.J.S acknowledges the hospitality of the
Department of Physics, Arizona State University, where much of
this work was completed. He also thanks S. V. Dmitriev for helpful
correspondence. A.J.S was supported by Grant NSF-DMR-0906491, and M.S.
was supported by JSPS-Grant-in-Aid for Scientific Research No.\ 25400394.
\end{acknowledgments}


\begin{thebibliography}{99}

\bibitem{Dol86} A. S. Dolgov, Fizika Tverdogo Tela {\bf 28} 1641 
[Sov. Phys. Solid State {\bf 28}, 907 (1986)].

\bibitem{ST88} A. Sievers and S. Takeno, Phys. Rev. Lett. {\bf 61}, 970
(1988).

\bibitem{Page90} J. B. Page, Phys. Rev. B {\bf 41}, 7835 (1990).

\bibitem{FW98} S. Flach and C. R. Willis, Phys. Repts. {\bf 295}, 182
(1998).

\bibitem{LS99} R. Lai and A. J. Sievers, Phys. Repts. {\bf 314}, 147
(1999).

\bibitem {CFK04}D. K. Campbell, S. Flach, and Y. S. Kivshar, Physics
Today {\bf 57}, 43 (2004).

\bibitem{FG08}  S. Flach and A. Gorbach, Phys. Repts. {\bf 467}, 1 (2008).

\bibitem{SS08} M. Sato and A. J. Sievers, Nature (London) 
{\bf 432}, 486 (2004).

\bibitem{WSS05} J. P. Wrubel, M. Sato, and A. J. Sievers,
Phys. Rev. Lett. {\bf 95}, 264101 (2005).

\bibitem{ETS08} L. Q. English, R. B. Thakur, and R. Stearrett,
Phys. Rev. E {\bf 77}, 066601 (2008).

\bibitem{EPS10} L. Q. English, F. Palmero, A. J. Sievers,
P. G. Kevrekidis, and D. H. Barnak, Phys. Rev. E {\bf 81}, 046605 (2010).

\bibitem{SHS06} M. Sato, B. E. Hubbard, and A. J. Sievers,
Rev. Mod. Phys. {\bf 78}, 137 (2006).

\bibitem{Sato11} M. Sato, S. Imai, N. Fujita, S. Nishimura, Y. Takao,
Y. Sada, B. E. Hubbard, B. Ilic, and A. J. Sievers, Phys. Rev. Lett. {\bf
107}, 234101 (2011).

\bibitem{EF05} M. Eleftheriou and S. Flach, Physica D {\bf 202}, 142 {2005}.

\bibitem{Manley09}M. E. Manley, A. J. Sievers, J. W. Lynn, S. A. Kiselev,
N. I. Agladze, Y. Chen, A. Llobet, and A. Alatas, Phys. Rev. B {\bf 79},
134304 (2009).

\bibitem{MA11} M. E. Manley, D. L. Abernathy, N. I. Agladze, and
A. J. Sievers, Nat. Sci. Repts {\bf 1}, 10 (2011).

\bibitem{DMCw} L. Z. Khadeeva and S. V. Dmitriev, Phys. Rev. B {\bf 81},
214306 (2010); {\bf 84}, 144304 (2011); A. A. Kistanov and S. V. Dmitriev, 
Fizika Tverdogo Tela {\bf 54}, 1545 (2012) [Phys. Solid State, {\bf 54}, 
1648 (2012)].

\bibitem{Kempa13} M. Kempa, P. Ondrejkovic, P. Bourges, J. Ollivier,
S. Rols, J. Kulda, S. Margueron, J. Hlinka, J. Phys.: Condens. Matter
{\bf 25}, 055403 (2013).

\bibitem{AM76} N. W. Ashcroft and N. D. Mermin, {\it Solid State Physics}
(Saunders College, Philadelphia, 1976).

\bibitem{BSW84} H. Bilz, D. Strauch, and R. K. Wehner, {\it Vibrational
Infrared and Raman Spectra of Non-Metals}, Handbuch der Physik, Vol.\
XXV (Springer-Verlag, Berlin, 1984).

\bibitem{KS97} S. A. Kiselev and A. J. Sievers, Phys. Rev. B {\bf 55},
5755 (1977).

\bibitem{SP95} A. J. Sievers and J. B. Page, in {\it Dynamical Properties
of Solids}, edited by G. K. Horton and A. A. Maradudin (North Holland,
Amsterdam, 1995), Vol. 7, p. 137.

\bibitem{Bilz79} See, for instance, H. Bilz and W. Kress, {\it Phonon
Dispersion Relations in Insulators} (Springer, Berlin, 1979).

\bibitem{RP00} T. R\"ossler and J. B. Page, Phys. Rev. B {\bf 62}, 11460
(2000).

\bibitem{KSC98} S. A. Kiselev, A. J. Sievers, G. V. Chester, Physica D:
Nonlinear Phenomena {\bf 123}, 393 (1998).

\bibitem{symmetryremark} We note that it is the higher point symmetry
of the fcc lattice relative to the zincblende lattice which suppresses
the anharmonic contribution to the vibration and forces such a large
amplitude to produce an ILM near the middle of the gap.

\bibitem{KT70} K. Tharmalingam, J. Phys. C:Solid St. Phys. {\bf 3},
1856 (1970) 

\bibitem{RP95} T. R\"ossler and J. B. Page, Phys. Lett. A {\bf 204},
418 (1995); Physica B {\bf 219-220}, 387 (1996).

\end{thebibliography}
\end{document}